\documentclass[11pt]{article}
\usepackage[utf8]{inputenc}
\usepackage[natbibapa]{apacite}
\usepackage{amsmath}
\usepackage{graphicx}
\usepackage{listings}
\usepackage{booktabs}
\usepackage{makecell}
\usepackage{lscape}
\usepackage{rotating}
\usepackage{lipsum}
\usepackage{lmodern}
\usepackage{dramatist}
\usepackage{mdframed}
\usepackage{newfloat}

\usepackage{report}

\DeclareFloatingEnvironment[
fileext=lot,
listname={List of Transcripts},
name=Transcript,
placement=tbp,
]{transcript}
\title{A Conceptual Framework for Modeling Team Adaptation in Cooperative Games Through Ludic Knowledge}
\shorttitle{A Shorter Title}

\author[1]{Caleb Vatral}
\affil[1]{Department of Computer Science, Tennessee State University}

\runningAuthor{Vatral}

\date{\today}

\begin{document}

\maketitle

\begin{abstract}
With the increasing importance of teamwork skills for modern workplaces, development of teamwork training programs has received substantial attention. Game-based teamwork training is one promising approach that is engaging, cost-effective, and well-suited to increasingly decentralized workplaces. However, design of effective game-based teamwork training requires understanding how a game elicits specific desired teamwork behaviors. Significant progress has been made in characterizing these relationships. However, despite its critical importance, little work has examined how a game's design influences team adaptability behaviors. 
This paper presents a preliminary framework for analyzing adaptability in cooperative games by conceptualizing adaptive stimuli as retrieval or disruption of players' ludic knowledge. We illustrate this framework through a qualitative case study that applies interaction analysis methods to gameplay videos of a Overcooked!, a cooperative cooking game. We examined instances where game events led to players altering their behavior and connected the game's design features that resulted in each event with three adaptive stimulus cue categories. Although exploratory and limited to a small case study of a single game, the proposed framework is grounded in established theories across teamwork research and game studies, and it offers an initial vocabulary for describing how cooperative games can be designed to create demands for team adaptation. With this continued development, the framework may provide an analytic tool to help inform the design and evaluation of purpose-built game-based teamwork training environments.

\end{abstract}

 
\section{Introduction}
Within modern workplaces, teamwork has become an increasingly critical component of success \citep{hosseinioun2025skill,HR2015}. 
One method of training these teamwork skills that has become increasingly promising is the play of cooperative digital video games. Games offer a highly engaging environment that is cost-effective and resilient to the challenges created by increasingly decentralized workplaces \citep{Chinyuku2025,Nadeem2023-ps,scacchi2009game}. A growing body of literature has begun to demonstrate specific links between game design elements and various teamwork behaviors \citep{Hmlinen2018,harris2019asymmetry,toups2014framework,Emmerich2017}. However, despite longstanding recognition as a critical component of effective teamwork \citep{salas2005Big5}, comparatively little work has focused on adaptability.

In this work, we a introduce preliminary conceptual framework for analyzing adaptability behaviors in cooperative games by synthesizing established models of adaptive stimuli and ludic knowledge. We argue that, in a game context, adaptive stimuli can be modeled as events that either activate or disrupt players' ludic knowledge. 
We illustrate this framework through a qualitative case study of a team playing through World 1 of Overcooked! \citep{Overcooked}, a cooperative cooking game. Using interaction analysis methods \citep{jordan1995interaction}, we examined player actions and communication to find instances where players encountered task demands that altered their behavior patterns. We identified the game's design features that resulted in each event and grouped these according to their adaptive stimulus cue type. 

Together, this model provides an initial vocabulary for describing how cooperative games can systematically create demands for team adaptation --- a first step toward training adaptability through games. Although exploratory and limited to a small case study, the proposed framework is grounded in established theories across teamwork research and game studies, and it offers a foundation for future work to apply and expand across a broader range of cooperative games. With this continued development, the framework may provide an analytic tool to help inform the design and evaluation of purpose-built game-based teamwork training environments. 

Through this analysis, this paper makes the following research contributions:
\begin{enumerate}
    \item \textbf{Theoretical:} Development of a preliminary conceptual framework to describe adaptability in cooperative games by synthesizing established theories of ludic knowledge and adaptive stimuli.
    \item \textbf{Empirical:} Qualitative interaction analysis of ecologically valid gameplay footage data that illustrates the use of the conceptual framework to characterize game events as activation or disruption of team ludic knowledge.
\end{enumerate}

\section{Adaptive Stimulus Cues}
Adaptability is the ability of a team to alter its course of action or team structures in response to information gathered from the environment. Burke et al.'s seminal model (\citeyear{burke2006understanding}) describes team adaptability in a four-phase cycle: situation assessment, plan formulation, plan execution, and team learning.
This cycle of adaptation begins with the \textit{adaptive stimulus}: a specific event or evolving situation that prompts a team to change their behaviors or structure in some way \citep{christian2017team}. 
There are a number of different factors that characterize an adaptive stimulus, but one of the most critical is the stimulus's \textit{cue}. The stimulus cue is the specific observable events and environmental details that are caused by the stimulus. Detecting the stimulus cue activates relevant schemata in long term memory, which enables continued situation assessment \citep{endsley2017toward,burke2006understanding}. 

Within a game context, stimulus cues are most often contextually related to in-game elements, events, and mechanics. Thus, the specific activated schemata will be a part of the team's shared \textit{ludic knowledge}. As conceptualized by Howell et al. (\citeyear{howell2014disrupting}), ludic knowledge is the specific knowledge that shapes players' expectations of a game. This is constructed through a combination a player's direct experiences with this game (intraludic knowledge), their experiences with other games (interludic and transludic knowledge), and their experiences outside games (extraludic knowledge). Based on retrieval of this knowledge, players make decisions and perform actions in the game world to accomplish their goals. 

Building on this model of ludic knowledge, we argue that schemata activated by in-game adaptive cues can take two possible forms: retrieval and disruption. In \textit{retrieval}, newly observed information related to the cue is compatible with existing schemata, and thus, the knowledge required to perform an adaptation can be recalled and utilized without substantial change. In \textit{disruption}, the newly observed information is incompatible with existing schemata, and thus requires the construction of new knowledge, either through restructuring existing schemata or creating completely new schemata.

\begin{figure}[t]
    \centering
    \includegraphics[width=0.9\linewidth]{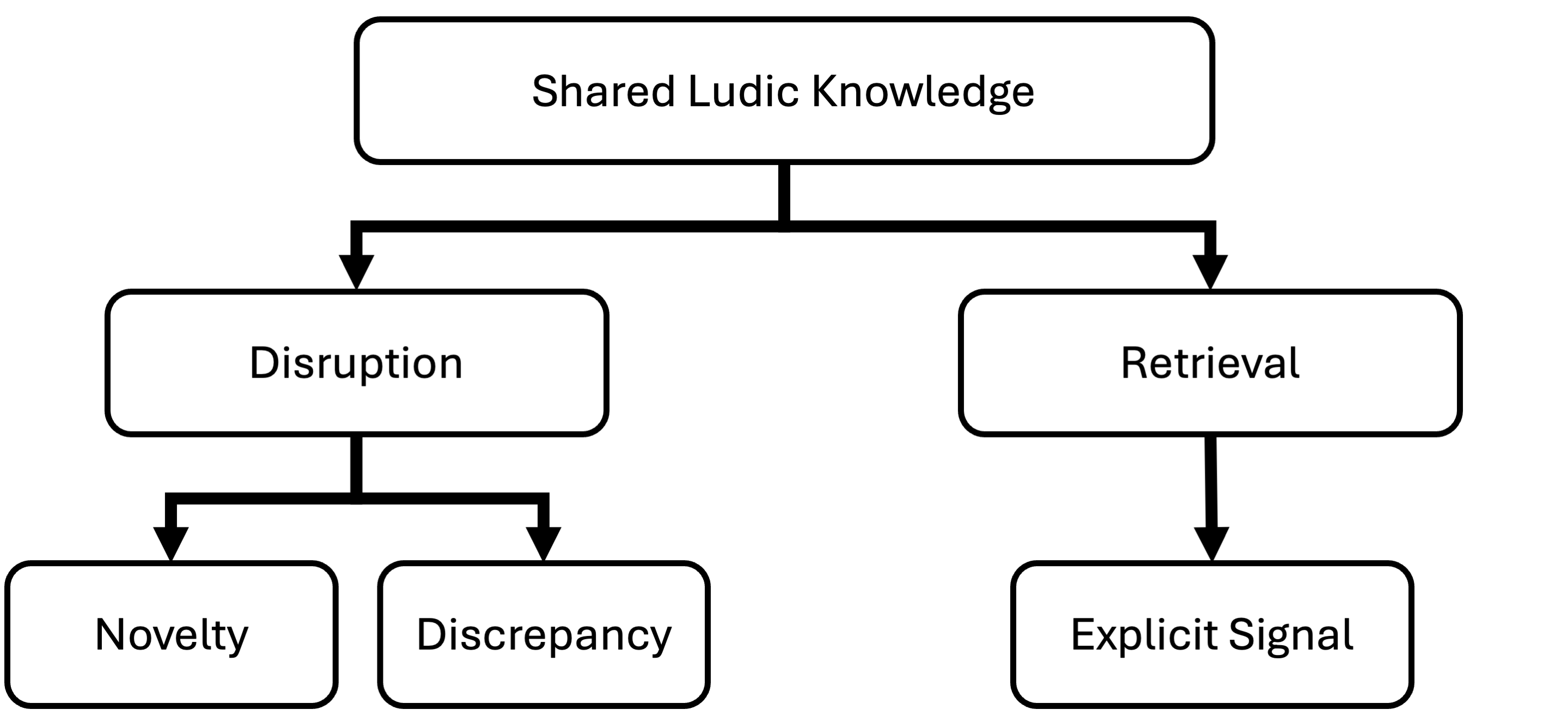}
    \caption{Conceptual model of adaptive stimulus cues in game environments.}
    \label{fig:ConceptualModel}
\end{figure}

Burke et al. (\citeyear{burke2006understanding}) described how Louis and Sutton's (\citeyear{louis1991switching}) model of cognitive modes can be used to understand adaptive stimulus cues. Since adaptation generally requires active thinking, the three conditions for changes from automatic processing to conscious engagement can be used as a model to describe situations in which teams are more likely to detect an adaptive stimulus. Building on this model offers an approach to further categorize our concept of disruption-based and retrieval-based adaptation cues. This conceptual model is illustrated in Figure \ref{fig:ConceptualModel}. 

Disruption-based cues are broken down into two explicit types of disruption from Louis and Sutton's model. First, \textit{novelty}-based disruptions result from players experiencing a completely new game element, event, or mechanic. Because it is the first time the players encounter it, new ludic knowledge needs to be generated to accommodate the experience. Second, \textit{discrepancy}-based disruptions result from the game environment reacting differently than players' expectations. These incorrect expectations are based on inaccurate or incomplete existing ludic knowledge, thus requiring updating of relevant schemata. In both cases of disruption-based cues, new ludic knowledge is constructed. Retrieval-based cues are highly related to Louis and Sutton's third condition, deliberate requests. In this case, some \textit{explicit signal} draws attention to the adaptive stimuli, which causes players to recognize the situation. However, different from disruption, in this case the situation is compatible with retrieved ludic knowledge, and thus players can apply the retrieved knowledge to respond to the stimuli without substantial modification.

\section{Methods}

To illustrate the application of our conceptual model, we performed a case study of one team of four people playing through World 1 of \textit{Overcooked!} \citep{Overcooked}. \textit{Overcooked!} is a cooperative cooking game where players work together to prepare recipes through various simplified cooking actions. However, the kitchen environments are designed with various hazards to impede progress and require player coordination. Since many of the hazards change over time within a given level, the game is well suited to studying adaptation.

Data was sourced from YouTube video recordings of gameplay and player voice. This type of secondary “found” data \citep{ang2013data} captures gameplay in a naturally occurring setting, thereby improving the potential ecological validity. We applied interaction analysis methods \citep{jordan1995interaction} to study the gameplay video. Candidate interactional sequences were selected when they contained observable evidence of team adaptation: an identifiable in-game event followed by a clearly discernible change in the team's behavior. Episodes were examined iteratively in both individual and group analysis sessions, focusing on characteristics of the adaptive trigger events, how teams recognized these events, and how the team's responses evidenced their ludic knowledge. The study was approved by the Tennessee State University Institutional Review Board under protocol HS-2026-5004.

\section{Results}

In this section, we will showcase three examples from the study that demonstrate the three adaptive stimulus cue types described in our conceptual model. These examples are a subset of the total corpus that was analyzed, and they were chosen to be illustrative of the broader patterns of adaptive stimulus cue types in the data. 

\subsection{Novelty-Based Cues}
The first type of adaptive stimulus cue is novelty-based cues. Falling under the boarder umbrella of ludic knowledge disruptions, novelty-based cues occur when the players experience a completely new element of the game for which they do not have explicit intra-ludic knowledge. This was the most commonly observed cue type in our data, which is sensible as it is very common practice in game design to introduce new mechanics and elements throughout the game.

\begin{figure}[t]
    \centering
    \includegraphics[width=\linewidth]{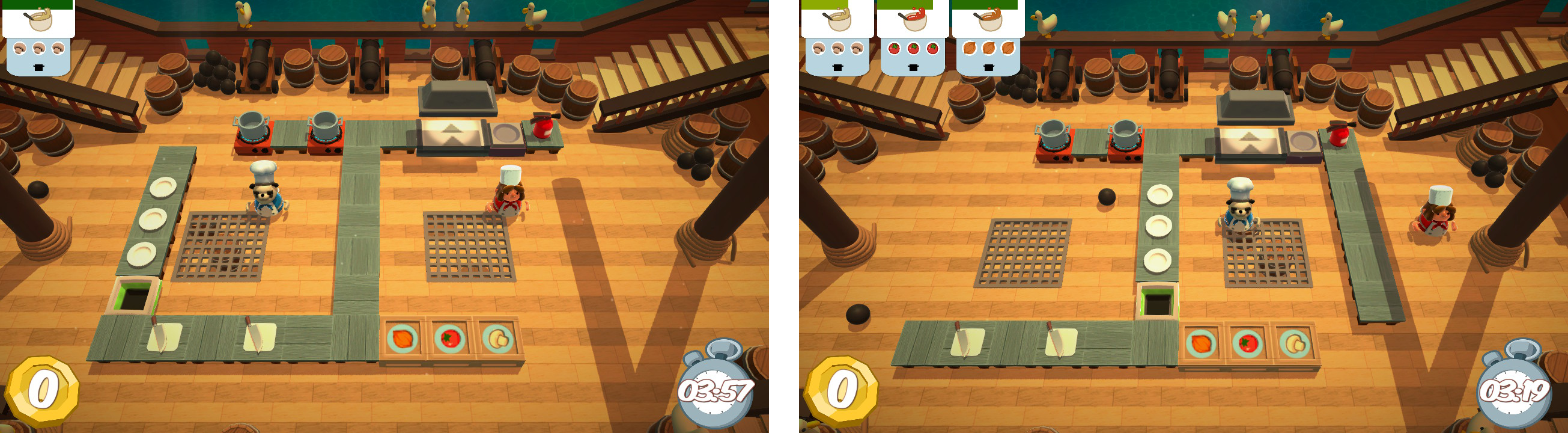}
    \caption{Level 1-3 from \textit{Overcooked!}, in which countertops move from the left and center of the level (left image) to the center and right of the level (right image) and vice versa.}
    \label{fig:Level3}
\end{figure}

In \textit{Overcooked!}, one clear example of novelty occurs in Level 1-3. In this level, the kitchen is placed on the deck of a pirate ship. Starting out, the level appears to be functionally similar to the previous levels, having players chop vegetables and cook them in pots to make soup. In fact, the team actually sees so many apparent similarities to previous levels that, right before the level begins, they discuss how the strategies they developed in previous levels (e.g., division of roles) could be carried over and adapted to this level. However, 30 seconds into the level, the ship tilts, causing some of the countertops to move positions. The level and this layout change is illustrated in Figure \ref{fig:Level3}. This is the first instance of a dynamically changing level layout seen in the game, so it is a novel situation for the players. Their surprise is very evident in their immediate reactions which contain exclamations, questions, and even evidence of frustration, as shown in Transcript \ref{transcript:Novelty}.

\begin{transcript}
    \begin{mdframed}
    \begin{drama}
      \Character{}{blank}
      \Character{Player 1}{one}
      \Character{Player 2}{two}
      \Character{Player 3}{three}
      \Character{Player 4}{four}
        \blankspeaks \direct{Ship tilts and countertops begin to move and push the players.}
        \onespeaks \direct{Simultaneously.} Oh! Oh my God!
        \twospeaks \direct{Simultaneously.} Oooh!
        \threespeaks \direct{Simultaneously.} Oh my God!
        \fourspeaks \direct{Simultaneously.} Ahh!
        \threespeaks Why would it do that?
        \onespeaks Onions. Need onions.
        \fourspeaks Need onions.
        \twospeaks Onions!
        \onespeaks \direct{Struggling to move around the countertops in their new positions.} Oh shit. Alright.
        \fourspeaks Onions coming up.
        \onespeaks \direct{Attempting to find a path to the pots.} Dammit. Got to go around.
    \end{drama}
    \end{mdframed}
    \caption{Players immediate reactions to the novel layout change in Level 1-3.}
    \label{transcript:Novelty}
\end{transcript}

The players initially adopt a role distribution strategy similar to what had been successful in previous levels: some players focused on getting ingredients, some players focused on chopping ingredients, and some players focused on cooking and plating. This causes them to spread out and position themselves around the level according to their role. However, when the level layout shifts, the players are pushed and paths are blocked, meaning they can no longer immediately access the resources needed for their assigned role. This causes the players to scramble and try to get back to their pre-negotiated roles, resulting in slowed progress and a pot almost catching on fire. The novel situation forced the team to make quick decisions about how to adapt, which in this case, was accomplished by shifting physical positions but not role assignments. 

As the level continued and this layout shift repeated, the team's adaptive response tended to become smoother. Players became familiar with both layouts and took quicker paths to transition between them. In one instance, a player temporarily stepped into another player's role to assist during the transition. This progression demonstrates the team's developing ludic knowledge which transitions these layout change events from novelty-based cues that require knowledge construction to retrieval-based cues that activate existing intra-ludic knowledge. This is analogous to the final learning phase often included in multiphase models of team adaptation \citep{burke2006understanding,pearsall2025stimulus}.

\subsection{Discrepancy-Based Cues}
The second type of adaptive stimulus cue, also under the umbrella of disruption, is discrepancy-based cues. Discrepancy occurs when the players' observations are inconsistent with their established ludic knowledge. In other words, something in the game environment is different from players' expectations. This was the least commonly observed cue type in our data, as it was difficult to find clear evidence of these cases since players often do not explicitly communicate their expectations. We hypothesize that these discrepancies actually occur quite often, but are often minor enough that players simply adjust their behaviors without explicitly communicating it. While we could make reasonable assumptions about players' mental models to identify more cases, in this work we follow common guidelines for interaction analysis to stay explicitly grounded in observable behaviors.  

Nonetheless, one very clear example of discrepancy occurred in the gameplay footage of Level 1-3 surrounding the dish washing task. Before the level begins, the team negotiates planned role assignments, with Player 3 volunteering to focus on washing and readying plates, as shown in Transcript \ref{transcript:Discrepency}. This role is a sensible assignment under the team's current ludic knowledge, as both of the previous two levels required dish washing between each use of a plate. However, this ludic knowledge is incomplete, as not all levels in \textit{Overcooked!} require dish washing. In some levels, such as Level 1-3, used plates are returned to the kitchen already clean and thus there is no sink in the kitchen. This causes a discrepancy cue, which is identified by the players through the sink's absence. This absence is noticed almost immediately by Player 3, who searches for the sink at the start of the level. However, Player 1 mistakenly identifies the trash can as the sink, so there is no immediate adaptation or knowledge revision to clear the discrepancy. Some time later, after the first used plate is returned to the kitchen, the discrepancy appears again. Player 3 picks up the returned plate and attempts to use the trash can, believing it is the sink based on Player 1's previous comments. After this action fails, the team has a short discussion and ultimately concludes that dish washing is not required for this level. This conclusion updates the team's ludic knowledge and causes Player 3 to quickly adapt their behaviors accordingly. Since there are less responsibilities related to readying the plates than anticipated, Player 3 switches to a combination of both plate preparation and assisting with ingredient gathering.

\begin{transcript}
    \begin{mdframed}
    \begin{drama}
      \Character{}{blank}
      \Character{Player 1}{one}
      \Character{Player 2}{two}
      \Character{Player 3}{three}
      \Character{Player 4}{four}
        \blankspeaks \direct{Discussion before Level 1-3 begins.}
        \twospeaks Okay, before we start, what's your job?
        \threespeaks I'll do plates.
        \onespeaks You're gonna do plates?
        \threespeaks I'll clean them. I'll provide them.
        \blankspeaks \direct{Continued discussion about role assignments.}
        \blankspeaks \ldots
        \blankspeaks \direct{Discussion shortly after the level begins.}
        \threespeaks Where's the fucking sink? Do y'all not see a sink?
        \onespeaks Its, its right there on the left. \direct{Mistakenly identifying the trash can.}
        \threespeaks \direct{Walking over to the trash can.} Is this it?
        \blankspeaks \direct{Players continue with their assigned tasks.}
        \blankspeaks \direct{\three moves the plates to sit between the pots and the delivery window.}
        \blankspeaks \ldots
        \blankspeaks \direct{Some time later, after the first order has been delivered and the used plate has just been returned to the kitchen.}
        \blankspeaks \direct{\three picks up the returned plate, which is already clean, and walks over to the previously identified trash can.}
        \threespeaks There's no?
        \twospeaks What do you need?
        \threespeaks There's no sink.
        \fourspeaks There is a sink. Its in the middle now. \direct{Referring to the trash can identified earlier, now in a new position from the level layout change.}
        \threespeaks That's not a sink!
        \blankspeaks \direct{Discussion continues until players conclude that dish washing is not required.}
    \end{drama}
    \end{mdframed}
    \caption{Group planning around dish washing and the subsequent discrepancy in Level 1-3.}
    \label{transcript:Discrepency}
\end{transcript}

\subsection{Explicit Signal-Based Cues}
The last type of adaptive stimulus cue is explicit signals, a form of retrieval-based cue. These signals draw player attention to the changing circumstances that require adaptation. Within \textit{Overcooked!}, there are a variety of these signals with various degrees of salience. For example, at the top of the screen, a set of UI elements display the current orders to be cooked and a timer for how long the players have left to make each order. Arguably, this timer acts as an explicit signal on its own, but becomes even more salient when the remaining time gets low and the UI element begins flashing red.

An even more clear example of these types of explicit signals in \textit{Overcooked!} is the cooking timer. When players place an ingredient in a pot or pan to cook it, a UI element appears near the stove and indicates the remaining cooking time. When the cooking completes, the timer disappears and a small animation plays with a green check mark. This animation and UI change act as a signal that a player needs to pause other tasks to remove the food from the burner. However, on their own, these signals are easy to miss, as occurred in Level 1-4 in our data. 

At the start of this sequence, Player 1 places a burger in one of the pans on a burner to cook. By this point in the game, the team has learned that scoring well generally requires them to work on other actions while the food cooks, not just stand and wait. So, after placing the burger to cook, Player 1 walks away and begins chopping vegetables. Busy with other tasks, the players do not notice that the burger has completed cooking, missing the initial checkmark animation signal. However, if a cooked dish stays on the burner too long, then the game initiates a secondary and more salient signal by flashing a red warning symbol and an audible beeping. If the players were to ignore this new signal for long enough, then the dish will catch on fire, causing a performance degradation. However, in this case, the players all immediately respond to the warning beep and scramble to reach the burner. The warning beep acts as an explicit signal that activates the team's ludic knowledge about the consequences of burned food. This knowledge prompts immediate adaptation to pause other ongoing tasks to instead focus on removing the food from the burner.

It is also important to note here that activation of existing ludic knowledge during retrieval-based cues does not necessarily mean that the subsequent adaptations will be optimal or even productive. In this case with Level 1-4, the team members all clearly recognized that the warning beep signal indicated a need to adapt and remove the food, but there was very little coordination of individual responses. This led to issues with players getting in each other's way as they all tried to fix the underlying issue. The explicit signal activated ludic knowledge for what needed to change, but not the method for coordinated execution.

\section{Discussion and Conclusion}

\subsection{Implications for Design}
The conceptual model developed and illustrated in this work suggests several implications for the design of cooperative games intended to elicit adaptability. Novelty offers a method for promoting adaptation that is well aligned with the incremental addition of mechanics that is already a common game design pattern. However, this should not be taken as a suggestion that every encounter should be novel. Not only would this be burdensome for designers, but since each cue type affords different cognitive and interactional demands, focusing only on one would also risk minimizing skills associated with the other two. For discrepancy cues, designers should seek to deliberately construct situations in which previously successful strategies become conditionally inappropriate. This forces players to practice knowledge revision and reorganization, which may be a particularly important skill for environments where information is limited or changing. For explicit signals, our findings indicate that these signals draw attention to changing task demands proportional to the signal salience. However, recognizing the need to adapt from such a signal does not guarantee effective coordinated execution. Thus, designers may want to consider adding scaffolding elements to guide development of coordination strategies when ludic knowledge is being applied for adaption. This may be particularly important when designing game-based teamwork training, where ensuring that adaptations are productive is the end goal.

\subsection{Limitations and Future Work}
While the results of this study are promising, it had several limitations that constrain its conclusions. Most importantly, the empirical analysis was highly limited in sample size and diversity. While this allowed for highly detailed analysis to illustrate the application of the conceptual model, researchers should be careful not to overgeneralize our results. Future work should apply our conceptual framework across a broader sample of cooperative games to determine whether the three cue types identified here are sufficiently general or whether additional categories are needed. Another limitation surrounds the use of found gameplay footage, which, while benefiting potential ecological validity, limited our access to players' internal expectations and knowledge. This was particularly consequential for identifying discrepancy-based cues. As noted in the results, players do not always explicitly communicate their expectations and mental models, meaning our analysis likely underestimates the frequency and variety of discrepancy-based adaptation. More controlled laboratory studies with explicit elicitation techniques, such as think-aloud protocols, may help close this gap. Finally, future work should extend the framework from recognition of adaptive stimuli to the quality of adaptive responses, which will be particularly important to enable the framework to be used to assist design of game-based teamwork training.

\section*{Acknowledgments}
The author would like to thank the members of the Virtual Interaction Analysis Lab, who generously participated in the group analysis sessions for this work. Their insights greatly strengthened the analysis and results. This work was supported in part by the UJIMA Project, which is funded by the National Science Foundation (Award \#2430323).

\bibliographystyle{agsm}
\bibliography{references}

\end{document}